\documentstyle[11pt]{article}
\newcommand{\be}{\begin{eqnarray}}

\newcommand{\ee}{\end{eqnarray}}

\title{
	\begin{flushright}
	{\normalsize TPI--MINN--93--44/T \\
	NUC--MINN--93--24/T \\
        HEP--UMN--TH--1220/93 \\
	August 1993 \\}
	\end{flushright}
\bf Computing Quark and Gluon Distribution Functions for
Very Large Nuclei }
\author{
	Larry McLerran and Raju Venugopalan \\
	{\small\it School of Physics and Astronomy,
 	University of Minnesota, Minneapolis, MN 55455} \\
	 }
\date{}

\begin{document}

\maketitle

\begin{center}
{\bf Abstract}\\
\end{center}
We argue that the distribution functions for quarks and gluons are computable
at small x for sufficiently large nuclei, perhaps larger than can be physically
realized. For such nuclei, we argue that weak coupling methods may be used. We
show that the computation of the distribution functions can be recast as a many
body problem with a modified propagator, a coupling constant which depends on
the multiplicity of particles per unit rapidity per unit area, and for
non-abelian gauge theories, some extra media dependent vertices.  We explicitly
compute the distribution function for gluons to lowest order, and argue how
they may be computed in higher order.

\vfill \eject

\section{Introduction}

The problem of computing the distribution functions for quarks and gluons in
hadrons is an old and difficult problem.  Issues such as Bjorken scaling were
greatly clarified by using light cone Hamiltonian methods~{\cite{soper} -
\cite{brodsky}}. There has been much progress recently in applying light cone
Hamiltonian methods together with non-perturbative methods gleaned from lattice
gauge theory technology to compute these distribution
functions~\cite{lclattice}. It has nevertheless been believed that the
computation of these distribution functions is intrinsically non-perturbative.

On the other hand, for a very large nucleus or at very small Bjorken x, it is
known that the density of quarks and gluons per unit area per unit rapidity
\be
	\rho = {1 \over {\pi R^2}} {{dN} \over {dy}}
\ee
is large.  For nuclei, we expect that this density scales as $A^{1/3}$, or
perhaps some larger power of $A$, so that for some sufficiently large $A$,
$\rho >> \Lambda^2_{QCD}$. Even if $\rho$ is not sufficiently large for
realistic nuclei, we certainly can imagine nuclei where this would be true. In
any case, when $\rho >> \Lambda^2_{QCD}$, we expect that the coupling constant
for strong interactions is weak, and weak coupling methods should be valid for
computing the distribution functions.~\cite{mueller}

In this paper, we will show how to set up the problem of computing the
distribution functions when the parton density $\rho$ is large. We will
explicitly compute the lowest order gluon distribution function at small
Bjorken $x$ for transverse momenta
\be
	\Lambda^2_{QCD} << \alpha_s \rho << k_t^2 << \rho
\ee
and find that in this kinematic region, the gluon distribution functions are of
the Weiszacker-Williams form
\be
	{1 \over {\pi R^2}} {{dN} \over {dx d^2k_t}} \sim
 {{\mu^2} \over {x k_t^2}}
\ee
where $\mu$ is a parameter which we shall compute which behaves as $A^{1/3}$.
In this theory, the dimensionful scale factor $\mu $ will set the scale of the
coupling constant.  All perturbation theory can be done in terms of $\alpha
(\mu )$, and if $\alpha (\mu ) << 1$ a weak coupling expansion is valid.  This
is equivalent to $\rho \sim \mu^2 >> \Lambda_{QCD}$.

We will also argue that the quark distribution functions are computable in this
kinematic region, and outline how to do the lowest order computation.  It may
be possible to extend the region of validity for computation of the
distribution functions to smaller values of $k_t$ by including non-perturbative
effects computable in weak coupling.  We also argue that the power dependence
of the distribution functions in Bjorken $x$ may be modified in higher orders
of perturbation theory.

The outline of this paper is as follows:  In the second section, we present a
brief review of the light cone quantization method, and our notation.  We also
argue that the valence quarks inside the nucleus may be replaced by static
quarks propagating along the light cone.  We set up the formalism which allows
us to compute the ground state properties of a large nucleus.  In the third
section, we do a simple analysis for QED and compute the infinite momentum
frame wavefunction for an electron to lowest non-trivial order in weak
coupling. We show that the photon distribution function is simply the
Weiszacker-Williams distribution of Lorentz boosted Coulomb photons.  In the
fourth section, we do the problem analogous to QED for large nuclei.  We argue
that for large enough nuclei for $k_t^2 << \rho$ where $\rho$ was defined
above, it is a good approximation to treat the local density of color charge
classically.  We argue that the local fluctuations in color charge may be
integrated out of the problem for computing ground state properties of the
system.  After performing this integration, we generate the distribution
functions for quarks and gluons.  In lowest order in perturbation theory, we
compute the distribution functions of gluons and show that they are of the
Weiszacker-Williams form.  We also show that there is an effective field theory
with propagators modified from their vacuum form which allows for the
systematic computation of corrections to the lowest order result.  The problem
of computing the distribution functions has therefore been converted into a
many body problem which for sufficiently large nuclei is a weak coupling
problem.  In the summary, we outline the computation of the quark distribution
functions.  We also show how non-perturbative, although presumably weak
coupling, effects modify the distribution functions at low $k_t$, $
\Lambda^2_{QCD} << k_t^2 << \alpha_s \rho$ We also argue that the $x_{Bj}$
dependence of the structure functions might be modified in higher orders in
perturbation theory.

\section{Notation and the Infinite Momentum Frame Hamiltonian}

Before turning to a detailed computation, we first review the
infinite momentum frame Hamiltonian method.  We will also
show how to include the effects of valence quarks which we will
treat as static sources moving along the light cone.

We begin with the action for the Yang-Mills field coupled to
fermions and also possibly some external current $J^{\mu}$,
\be
	S = \sum_i \overline \psi_i (\not \! P + M_i ) \psi_i
  +{1 \over 4} F_{\mu \nu} F^{\mu \nu} - J \cdot A
\ee
Here the Yang-Mills field strength is
\be
	F^{\mu \nu}_a = \partial^\mu A^\nu_a - \partial^\nu A^\mu_a
  -g f_{abc} A^\mu_bA^\nu_c,
\ee
the covariant momentum is
\be
	P = {1 \over i} \partial - g A,
\ee
where we use the notation that
\be
	A^\mu = A^\mu_a \tau^a
\ee
We are using the metric where $a \cdot b = a^i b^i - a^0 b^0$,
but with the ordinary convention for gamma matrices.

This Hamiltonian
may be re-expressed in terms of light cone variables by the
identification of components of vectors as
\be
	a_{\pm}  =  (a_0 \pm a_3 )/ \sqrt{2}
\ee
In these variables, the metric is $g_{+-} = -1 = g_{-+}$, the transverse
components of the metric are the unit matrix, and all other
components vanish.

To quantize along the light cone, as usual we first express the fermion fields
in terms of their light cone spinor components by introducing the
projection operators
\be
	\alpha^\pm = {1 \over \sqrt{2}} \gamma^0 \gamma^\pm
\ee
These are Hermitian projection operators so that we can write
\be
	\psi_\pm = \alpha^\pm \psi
\ee
where
\be
	\alpha^\pm \psi_\pm = \psi_\pm
\ee
and
\be
	\alpha^\pm \psi_\mp = 0
\ee
so that
\be
	\psi = \psi_{+} + \psi_{-}
\ee

We are going to be interested in the light cone Hamiltonian
$P^{-}$ which generates displacements in $x^+$.  When we write out
the Euler-Lagrange equations for the fermion fields, one
of the equations is a constraint equation for the fields
on a surface of fixed $x^+$,
\be
	\sqrt{2} P_{-} \psi_{-} = - \gamma^0 (\not \! P_t + M) \psi_{+}
\ee
This equation can be explicitly inverted in the light cone gauge
\be
	A_{-} = - A^{+} = 0
\ee
We find
\be
	\psi_{-} = {1 \over {\sqrt{2}P^+}} \gamma^0 (\not \!  P_t + M)
\psi_{+}
\ee
The fermion contribution to the action is therefore
\be
	S_F = -\psi_+^\dagger P^- \psi_+ + {1 \over 2} \psi_+^\dagger
(M- \not \! P_t) { 1 \over P^+ } (M + \not \! P_t ) \psi_+
\ee
where we have rescaled $\psi \rightarrow {1 \over 2^{1/4}} \psi$.
In terms of these variables, we see that $\psi_+^\dagger$ is the
light cone momentum canonically conjugate to $\psi_+$.

To analyze the vector contribution to the action, we first write explicitly
\be
	F^2 = F_t^2 - 4F_{k+} F_{k-} + 2F_{+-}F_{+-}
\ee
In light cone gauge, we have
\be
	F_{+-} = \partial_+ A_- - \partial_- A_+ - ig [ A_-,A_+ ] =
 -\partial_- A_+,
\ee
\be
	F_{k+} = \partial_k A_+ - \partial_+ A_k - ig [ A_k,A_+ ],
\ee
and
\be
	F_{k-} = E_k = -\partial_- A_k
\ee

The equations of motion for the vector field are
\be
	D_{\mu} F^{\mu \nu} = J^\nu
\ee
In particular, the equation for the $+$ component of the current is a
constraint equation for $A^-$ on a fixed $x^+$ surface,
\be
	-\partial_-^2 A^- = J^+_F + D_k E^k
\ee
where $J_F$ is the current generated by the fermion field
\be
	J_{Fa}^\mu = \overline \psi \gamma^\mu \tau_a \psi
\ee
plus whatever external currents there are in the system.  The piece
of the right hand side of this equation which is
\be
	 D_k E^k
\ee
can be thought of as the bosonic contribution to the $+$ component
of the charge density to which $A^-$ couples.

We can now write the bosonic part of the action as
\be
	{1 \over 4} F^2 = {1 \over 4} F_t^2 -  D_kE^k A_+ +
{1 \over 2 } (\partial_- A_+ )^2 - \partial_- A_k \partial_+A^k
\ee

We see that the momentum canonically conjugate to the field $A_k$
is the momentum $\Pi_k = -E_k$ where the index $k$ runs only over
transverse coordinates.  We can then express the action in terms of
momenta and coordinates as
\begin{eqnarray}
	S & = & { 1 \over 4} F_t^2 + {1 \over 2} \left( \rho_F + D_t \cdot E_t
\right) {1 \over {P^{+}}^2} \left( \rho_F + D_t \cdot E_t
\right) +  \nonumber \\
        & & {1 \over 2}
\psi^\dagger (M-\not \! P_t ) { 1 \over P^+ } (M+\not \! P_t) \psi
-i\psi^\dagger \partial_+ \psi + E_k \partial_+ A_k
\end{eqnarray}

We therefore see that the generator of $x^+$ transformations is
\begin{eqnarray}
	P^- & = & { 1 \over 4} F_t^2
+ {1 \over 2} \left( \rho_F + D_t \cdot E_t
\right) {1 \over {P^{+}}^2} \left( \rho_F + D_t \cdot E_t
\right) + \nonumber \\
   & &\psi^\dagger (M-\not \! P_t ) { 1 \over P^+ } (M+\not \! P_t) \psi
\end{eqnarray}
With the above identification of canonical momenta and coordinates,
we see that we can quantize the fermion and vector fields as
\be
	\psi_\alpha (x) = \int_{k^+ > 0} {{d^3k} \over {(2\pi)^3}}
\left( b_\alpha (k) e^{ikx} + d^\dagger_\alpha (k) e^{-ikx} \right)
\ee
and
\be
	A_i^a (x) = \int_{k^+ > 0} {{d^3k} \over {(2\pi )^3 \sqrt{2k^+}}}
\left( a_i^a (k) e^{ikx} + a^{a\dagger}_i (k) e^{-ikx} \right)
\ee
The commutation relations for the operators $a,b$ and $d$ are
\be
	[b_\alpha (\vec{k}) , b^\dagger_\beta (\vec{k'}) ]_+
         = [d_\alpha (\vec{k}) , d^\dagger_\beta (\vec{k'}) ]_+
        = (2\pi )^3 \delta^{(3)} (\vec{k} - \vec{k'} ) \delta_{\alpha \beta}
\ee
and
\be
	[a^i_a(\vec{k}), a^{j\dagger}_b(\vec{k'})] = (2\pi )^3 \delta^{(3)}
(\vec{k} - \vec{k'}) \delta^{ij} \delta_{ab}
\ee
with all other (anti)commutators vanishing.

The above review of the light cone quantization procedure will standardize
the notation for the following analysis.  We have seen that it is
possible to explicitly eliminate the constraints of gauge fixing in the
light cone gauge and get a light cone Hamiltonian expressed in
terms of the true dynamical degrees of freedom of the system.

Finally, we will discuss an essential part of our formalism: the reduction
of the valence quarks for the infinite momentum frame wavefunction
to static external sources of charge.  To do this, we first make the
assumption that in the end we will be applying our analysis
to weakly coupled systems.  In a theory like QED, this can be done
automatically since in this theory the coupling constant is always
weak, except possibly at very short distances.  In QCD, we will
have to restrict our attention to systems where the density of partons
per unit area is very large.

If the coupling is weak, then the dominant mechanism for producing
a cloud of partons around a fast moving valence particle is
by bremstrahlung or chains of particles bremstrahlunging
from bremstrahlung particles.  In this case,
for weak coupling, the typical momentum being transferred in the
bremstrahlung process is soft, and the valence parton does not lose
a large fraction of its momenta.  If this is the case, then
the valence parton typically moves with a velocity close to
light velocity with only very small transverse components generated
by coupling to multiparticle degrees of freedom.  The valence
parton is therefore a recoiless source of charge moving
along the lightcone.  It is well known that this approximation
describes the soft photon dressing of the electron in QED,
and has been used to study the infrared region of QED.

In QCD, we are claiming that the approximation should be valid for describing
the parton distributions generated by the valence parton
whenever the density of partons is sufficiently high that
weak coupling methods can be used.  This will however only work
for describing the parton cloud generated at small $x_F$.  In
the fragmentation region, we are sensitive to details of the
spatial distribution of quarks as generated by the nuclear
wavefunction.  For example, if we look at the distribution
of nucleons in the rest frame of a nucleus, we see the
extended nuclear structure.  If we however look in the
central region, the valence nucleons will appear to be
Lorentz contracted to a pancake of dimensions of
order $R_{nuc} / \gamma_{cm}$ where $R_{nuc}$ is the
nuclear radius, and $\gamma_{cm}$ is the gamma factor
for the nucleus as measured in the center of mass frame.
In the central region, we expect that the parton distributions
will be insensitive to the details of the distribution of partons
in the fragmentation region, that is, the valence partons will act
simply as a source of charge which is essentially a delta
function along the light cone.

In addition, maintaining the constraint that the momentum
transfer be small and the coupling weak is a little tricky in QCD.
We will require that the momentum transfer be small compared
to the total momentum of the valence partons, but large compared
to the QCD scale.  For this to be consistent, the low momentum
range of integration must be cutoff for small momentum
transfer for the range of dynamically important momenta.
Presumably this happens due to media effects caused by the high
density of partons.

Although the arguments given above are heuristic, we will
find that the solution we generate for the quark and gluon
distribution functions in the central region are
self-consistent with the assumption that the valence partons may
be treated as a delta function source of charge along the light
cone. Perhaps it might be possible at some point to relax the assumptions
about the fragmentation region distributions, but we do not at
this point know how to do it.

We therefore will treat the valence parton distributions
as recoiless sources of charge which are localized along the
light cone, that is, we take
\be
	J^+_a = Q_a(x^+ )  \delta (x^- )
                 \delta^{(2)} (\vec{x_t} )
\ee
where we have assumed the source is localized at $x^- = 0$ and
$\vec{x_t} = 0$.  (We could have placed the sources anywhere.)
The charge $Q_a$ is an operator which has the charge algebra
\be
	[Q_a,Q_b] = if_{abc}Q_c
\ee
and is in some representation of the Lie group corresponding
to the structure constant $f_{abc}$.  If the representation of the
Lie algebra is sufficiently large, we may be able to replace
the charge operator $Q_a$ by a c-number classical source.

The $x^+$ dependence of the source $Q_a$ is determined by a covariant
conservation law
\be
	\partial_+ Q_a = i f_{abc} A_b Q_c
\ee
This equation is necessary so that the equations of motion
for the vector fields are consistent,
that is we must have
\be
	0 = D_\nu D_\mu F^{\mu \nu} = D_\nu J^\nu
\ee
This equation would be immediately true if we took the current
for dynamical fermion fields.  We can take the
static limit for fermion fields and construct the current,
and the above equation will therefore be valid.

We can solve the equation for the time dependence of the charge
operator.  If we let
\begin{eqnarray}
	Q & = & \tau^a Q_a \\
        A_+ & = & = \tau^a A_{+a} \\
\end{eqnarray}
Further let the time ordered exponential be defined as
\be
	Te^{i\int_0^{x^+} dx^{+'} A_+(x^{+'},x^-,x_t )}
           = U(x^+,0)
\ee
where we are not writing out the dependence of $U$ upon
$x^-$ and $x_t$.
We therefore have that
\be
	Q(x^+ ) = U(x^+,0) Q(0) U(0,x^+)
\ee
Note that $U$ is a unitary matrix and that
\be
	U(0,x^+) = U^{-1}(x^+,0)
\ee
so that the time evolution of the charge operator
is just a rotation of the charge operator in charge space.

We emphasize here that the above equation for the time evolution
of the charge operator is true for classical as well as
quantum charge operators.  For abelian theories, the charge
is simply time independent.  In non-abelian theories, the
charge is specified by its initial value, but after this time
rotates in charge space.

\section{The Photon Distribution Function of an Electron}

The discussion of the previous section has been highly formal,
and largely a brief review of what is already known in the literature.
The introduction of static sources along the light cone in the
form presented above is perhaps less well known, and it is useful
to see how this formalism works in the case of QED.
Here we will compute the wavefunction for an electron moving along
the lightcone, and the photon distribution function of the electron.
It will not be too surprising that we find that the distribution
function is precisely that for the Weiszacker-Williams
distribution of photons generated by Lorentz boosting the Coulomb
field.

We will work to lowest non-trivial order
in $\alpha $ so that we can ignore $e^+e^-$ pair production.
In this approximation, we ignore dynamical fermions
in the light cone Hamiltonian.  The source term for the
valence electron is
\be
	\rho_e = e\delta (x^-) \delta^{(2)} (\vec{x_t} )
\ee
The light cone Hamiltonian is
\be
	P^- = \int d^3x \left( {1 \over 4} F_t^2
+ {1 \over 2} (\rho_e + \nabla_t \cdot E_t ) {1 \over P^{+2} }
(\rho_e + \nabla_t \cdot E_t ) \right)
\ee

The ground state for this light cone Hamiltonian is a coherent state
\be
	\mid \Psi > = C ~\exp\left( i \int d^3x A^{op} (x) \cdot E^{cl} (x)
                         \right) \mid 0 >
\ee
In this expression, the quantity $C$ is a normalization constant,
the field $A^{op}$ is the operator value of the transverse
component of the vector potential, and $E^{cl}$ is a c-number valued
classical electric field.

The reason ground state is a coherent
state is easily understood.  In the Weiszacker-Williams approximation, the
photon cloud is a classical distribution of electromagnetic
fields generated by Lorentz boosting the Coulomb field.
It is well known that coherent states describe classical
fields.

The expectation value of the light cone Hamiltonian in this state
is simply
\be
	P^- = \int d^3x {1 \over 2}
(\rho_e + \nabla_t \cdot E_t^{cl} ) {1 \over P^{+2} }
(\rho_e + \nabla_t \cdot E_t^{cl} )
\ee
where we have used that the exponential in the coherent state
simply shifts the electric field.
Extremizing the Hamiltonian with respect to the classical
electric field gives
\be
	\nabla_t \cdot E_{cl} = -\rho_e
\ee
so that the classical electric field may be taken as purely longitudinal
in the two dimensional transverse space.  This is consistent with ignoring
the contribution from $B^2$ which arises from transverse components.
(A more careful  treatment would have involved writing the Hamiltonian
in terms of transverse and longitudinal creation and annihilation
operators, constructing a coherent state in terms of transverse and
longitudinal fields, and showing that the minimum energy configuration
has zero transverse vector potential.)

An amusing feature of the QED analysis is that the eigenvalue
of the light cone Hamiltonian for the ground state vanishes.
This is because only a longitudinal field arises and hence
there is no contribution from the transverse magnetic field.
The light cone charge of the source is also identically
canceled by the light cone charge generated by the
transverse electric field.

We therefore have the classical electric field as
\be
	E_t^l (k) = ie {k^l_t \over k_t^2}
\ee
and the vector potential as
\be
	A_t^l (k) =  e {1 \over k^+}  {k^l_t \over k_t^2}
\ee
where our convention for Fourier transformation is
\be
	F(\vec{x},x^+) = \int~{{d^3k} \over {(2\pi )^3}} \overline F
(\vec{k},x^+)
\ee
Notice that the transverse electric field which we
have generated is precisely the
Weiszacker-Williams field generated by Lorentz boosting the
Coulomb field.  We could further reproduce the magnetic components
of the field if we were to compute $A_+$, and its associated field
strengths.

The photon distribution function is given by
\begin{eqnarray}
	F(k^+,k_t) & = & <a^\dagger_l (k) a^l (k) > \\
                   & = & {{2 e^2} \over {(2\pi)^3}} {1 \over {k^+ k_t^2}} \\
\end{eqnarray}
where $F(k^+, k_t) = dN/dk^+d^2k_t $
This can also be rewritten in terms of the Bjorken x variable as
\be
	F(x,k_t) = {\alpha \over \pi^2}  {1 \over {x k_t^2}}
\ee

The re-expression in terms of Bjorken x variables may here seem a bit
peculiar since nowhere has the longitudinal momentum of the source
been introduced.  It can be done because of the scaling property
of the distribution function, peculiar to the $1/k^+$ behavior.
If there was any other power of x, this could not be done.
We might therefore ask what would happen in the QCD case or
perhaps in QED if electron-positron pair creation was included.
Presumably the distribution function becomes modified to be something
like
\be
	F(x) \sim {1 \over x^{1+ C \alpha} }
\ee
so that higher orders generate terms which involve logarithms of x. These
logarithms of $k^+/P_{source}^+$ presumably arise in our formalism by a
sensitivity to a high momentum cutoff necessary to regularize the
$\delta$-function distributions for the external sources. Note that for a
delta-function source, there is a contribution from arbitrarily large momenta.
This is presumably cutoff at momenta of order that of the valence particle when
recoil is included. If this is the case, then we expect that we will be able to
compute the  dependence of the distribution functions on $x$ and $k_t$, up to
some overall constant. This constant cannot be computed without a better
knowledge of the dynamics of the fragmentation region, that is understanding
some of the relevant details of the valence particle recoil.  In any case, it
is remarkable that, to lowest order, the result is insensitive to such details.

\section{Ground State Properties in QCD in the Presence
of Sources}

In QCD, we shall be interested in computing the distribution functions
for quarks and gluons generated by some valence distribution
of quarks.  In order to be able to use weak coupling techniques,
we will have to require that the density of partons per unit
area
\be
	\rho = {1 \over {\pi R^2} } {{dN} \over {dy} }
\ee
is large
\be
	\rho >> \Lambda^2_{QCD}
\ee

The simplest problem to consider is that of the distribution functions
for a very large nucleus.  Although our results could be generalized
to a finite size nucleus, we will consider infinite nuclear matter
in the transverse direction with a uniform transverse density distribution.
We will assume that the nucleus is thick compared to a proton,
that is that the local density of baryon number per square fermi is
large.  Roughly speaking, the density of baryon number per square fermi
should be of order $A^{1/3}$ for an ordinary nucleus, and
the transverse extent should also be of order $A^{1/3}$.  We could also
presumably take the distribution functions we compute here as a function
of local baryon density per unit area and convolute them over
a density distribution for a nucleus in order to determine
realistic parton densities.  For realistic nuclei,
our weak coupling approximations are quite probably at the edge of
being valid.

We shall therefore imagine that the baryons which generate the nuclear
valence distribution are localized on the light cone but uniform
in transverse space.  Their charge will however in general be
fluctuating around zero from one transverse position in space to the
next.  We expect that the average color charge associated with the
valence distributions will be zero, and therefore the only way
to generate a non-zero source for charge is by fluctuations.
We will treat the nuclear valence quarks as static, that is
recoiless, sources of charge.

To compute ground state properties of our ensemble of charges,
we consider the ground state expectation value
\be
	Z = <0 \mid e^{iT P^-} \mid 0 >
\ee
where $T$ is a parameter which will be taken to be large
in the end.  The state $\mid 0>$ is the ground state in the
presence of a uniform density of sources of color charge.

If we take the limit that $T \rightarrow \infty$, we can generate
ground state expectation values by
\be
	Z = \lim_{T \rightarrow \infty}~ \mbox{Tr}~ e^{iTP^-}
\ee

The problem with evaluating the trace is that for quantized sources
of color charge, the sum over different values of the color charges
is difficult to evaluate.  However for a large nucleus, so long
as we resolve the system on a transverse size scale which is much larger
than a typical transverse quark separation, we expect that many
different quark charges will contribute to the valence charge
density.  A large number of charges corresponds to a high dimensional
representation of the color algebra, and therefore the sum
of the color charges of the quarks can be treated classically.

To be more specific, we look at a region of transverse extent where
there are a large number of quarks.  The total charge in this region
on the average will be zero.  There will however be fluctuations in the
color charge.  We would naively expect that the typical value
of the fluctuating color charge would be  of order $\sqrt{N}$
where $N$ is the number of quarks in this
region~{\cite{ehtamo}--\cite{eisenberg}}.  If $N$ is large, the typical
color charge in this region is large, and therefore may be treated
classically.

This may be analyzed in a quantitative manner using our formula for
$Z$ which generates ground state expectation values.  The point is that
in the trace, there is the sum over all possible color orientations
of the external source corresponding to the valence quark distributions.
Let us make a grid in the transverse space so that each grid has
many valence quarks in it.  This therefore restricts the validity of
our analysis to spatial resolutions with
\be
	d^2x >> 1/\rho_{valence} \sim A^{-1/3} fm^2
\ee
or
\be
	q_t^2 << (200~ MeV)^2~ A^{1/3}
\ee
For realistic large nuclei, this corresponds to $q_t \sim 1-2$ GeV
(recalling that in the center of the nucleus the effective
value of $A^{1/3}$ is larger than average), and in this
kinematic region, weak coupling methods are at best marginal.
In principle we can imagine nuclei with very large $A$,
so we can consider this as a theoretical laboratory.
The situation might also be improved if the density of partons
per unit area in the central region were to greatly exceed this and
that the coupling constant were smaller than naively expected from the
above consideration.

Now in our sum over states in the formula for $Z$, there is a large
number of states being generated for each transverse area.
We would like to determine the set of most probable configurations.
To do this, we need to construct the density of states for
such configurations.  We know that the maximum will be centered
around an average color charge of zero.  Since it is a maximum,
it must be true that it is quadratic in the charge density.
Therefore the contribution to Z is of the form
\be
	\exp\left(-{1 \over {2\mu^2}} \int d^2x_t \rho^2(x) \right)
\ee
Here $\rho^2 $ is the charge density squared measured in
units of $g$, that is with the factor of $g$ extracted, and
$\mu^2 $ is the average charge density squared per unit area
divided by $g$.

The above formula for the density of states is valid so long as
the average charge is of the order of the $\sqrt{N_0}$.  In this case,
higher order corrections to the Gaussian approximation for the density
of states go like $1/\sqrt{N_0}$.  Notice that with the
assumption that the number of particles in our box is of
order $N_0$, we have that $d^2x \rho^2 \sim 1$,
and higher order terms are correspondingly smaller in powers of
$N_0$ since $\rho \sim \rho_Q/\sqrt{N_0}$ where $\rho_Q$ is the
density of valence quarks per unit area.

To see how such a density of charge works,
we see that if we define
\be
	<A> = { {\int [d\rho ]~  A~
\exp\left(-{1 \over {2\mu^2}} \int d^2x_t \rho^2(x) \right)}
\over {\int [d\rho ]~
\exp\left(-{1 \over {2\mu^2}} \int d^2x_t \rho^2(x) \right)} }
\ee
then we have
\be
	<\rho (x) > = 0
\ee
as we have assumed and
\be
	<\rho (x) \rho (y) > = \mu^2 \delta^{(2)} (\vec{x} - \vec{y} )
\ee

We can determine the value of $\mu^2$ from elementary considerations.
The number of quarks per unit area in a nucleus is
\be
	n_q = 3 (A/\pi R^2) \sim 0.8~ A^{1/3} fm^{-2}
\ee
and the average charge squared of a quark is
\be
	<Q^2> = g^2 \sum_a \tau_a^2 = {4 \over 3} g^2
\ee
so that
\be
	\mu^2 = 1.1~A^{1/3} fm^{-2}
\ee

Now that we understand the density of charge states associated
with the external charge, we proceed to a path integral representation
for Z.  This is easily done in the standard way by introducing
states which are momentum and coordinate basis states for the fields.
The only tricky part is to recall that in integrating over the
external charges, we must remember that the external charges
must satisfy the extended current conservation law discussed in the
second section.  The result of some analysis is
that in light cone gauge
\be
	Z & = & \int~ [dA_t dA_+] [d\psi^\dagger d\psi] [d\rho]
 \nonumber \\
& & exp\left( iS +ig\int d^4x A_+(x)\delta (x^-)
\rho (x)  - {1 \over {2\mu^2}} \int d^2x_t \rho^2 (0,x_t) \right)
\ee
In the functional integral, we have that the path integral over the
fields is at all space-time points.  The path integral
over $\rho $ is at only a fixed value of $x^+$ which we choose to
be $x^+ = 0$, and we define $\rho $ at other values of $x^+$
by
\be
	\tau_a \rho^a (x^+) = U(x^+,0) \tau_a \rho^a (0) U(0,x^+)
\ee
where $U$ is the time ordered exponential of $A_+$ as discussed
in the second section.

To get an action expressed only in terms of the vector fields,
we must perform the functional integral over fluctuations in the
external charges.  This is straightforward to do because the
action is quadratic in the external charges.  The part of the action
involving $\rho$ is
\be
	-i S_\rho & = & - ig\int_{T/2}^{T/2} dx^+ \int d^2x_t A_+^a (x)
\rho^b (0;x_t) M^{ab}(x^+,0;x^-,x_t) \mid_{x^- = 0} \nonumber \\
            & &     +{1 \over {2\mu^2}} \int d^2x_t \rho^2 (0;x_t)
\ee
In this expression, the quantity $M^{ab}$ is
\be
	M^{ab} (x^+,0;x^-,x_t)
  =  2~ Tr \left( \tau^a U(x^+,0;x^-,x_t) \tau^b U^(0,x^+;
x^-,x_t) \right)
\ee
Notice that in this expression there is non-locality in the time variable
$x^+$, but locality in $x^-$ and $x_t$.
The matrix $M$ is Hermitian since
\be
	 M^{ab} (x^+,0;x^-,x_t) = M^{ba*}(0,x^+;x^-,x_t).
\ee

The integral over $\rho$ can be performed since the integral is Gaussian.
The value of $\rho $ is given by
\be
	\rho^b (0) = i g \mu^2 \int dx^+~ A^a_+ (x) M^{ab} (x^+,0;x^-,x_t)
\mid_{x^- = 0}
\ee
Upon integrating out the sources, we obtain that the contribution of
$S_\rho$ becomes
\be
	-iS_\rho & \rightarrow & g^2 {\mu^2 \over 2} \int dx^+dy^+ \int d^2x_t
{}~ A^a_+(x^+,x^-,x_t) A^b_+(y^+,x^-,x_t) M^{ac}(x^+,0;x^-,x_t) \nonumber \\
 & & M^{bc}(y^+,0;x^-,x_t)\mid_{x^- = 0}
\ee

In QED, the effect of integrating out the sources is simply to give
\be
	S_\rho \rightarrow g^2 {\mu^2 \over 2} \int dx^+dy^+d^2x
A_+(x^+,x^-,x_t)A_+(y^+,x^-,x_t)\mid_{x^- = 0}
\ee
that is, only the propagator is modified by the addition of an extra
term, which is formally of order $\mu^2 g^2$.
In QCD, the problem is more complicated.  In addition to the
modification of the propagator in order $g^2$, there are in addition
a non-polynomial and non-local set of vertices which contain arbitrary
powers of the field $A_+$.  These vertices will modify the
Feynman rules in orders beyond the lowest order formally in $g^2$.

We must ask what is the expansion parameter of the theory.
Later in this section, we will show that the distribution function
for the number of gluons per unit area is given in lowest order by
\be
      {{F(x,k_t) } \over {\pi R^2}} = g^2 \mu^2 {{2(N_c^2-1)} \over {(2\pi)^3}}
 {1 \over {k^+k_t^2}}
\ee
In general, when we are in the limit that the relevant momentum
scales are
\be
	g^2 \mu^2 << k^2 << \mu^2
\ee
we are in the low density small $g^2 \mu^2$ limit so that the
effects of the modified gluon propagator may be treated
perturbatively.  The other end of the limit, that $k^2 << \mu^2$
is just the statement that we are at sufficiently small
momenta that the individual quarks which are the source of
color charge cannot be individually resolved.

In the above kinematic limit, a treatment of our effective Lagrangian
to lowest order in weak coupling and lowest order in $\mu^2$ is
justified.  The higher orders in $g^2$ should be small so long as
$g^2$ is small due to the high density of partons.  This
may be only part of the story however.  As is known in QCD at finite
temperature, it may be true that at some order in perturbation theory
for the computation of physical observables, the weak coupling
expansion becomes essentially non-perturbative.  This could happen
here, and to know if it happens and what order it affects observables
is beyond the scope of the present work.

In addition to the above problems, the constraints of the renormalization
group equations must be satisfied.  This will give a non-trivial
dependence on coupling to the structure functions.  Beyond this
there may also be singularities which arise from approximating the source
of color charge as recoiless.  For example, if the structure
functions end up having a power of Bjorken $x$ other than
$1/x$, then a scale associated
with the longitudinal momentum of the sources will have to
come into the problem.  This will arise from regularizing the
recoiless charge distribution which for a delta function
involves integrating over all longitudinal momentum.  If we cut
this off at some value of $P^+$, then if the corrections to the
lowest order result for the distribution functions involve
logarithms of this cutoff scale, they can sum up to
something like
\be
	\sum_n  C^n\alpha_s^n log^n(p^+/k^+)/n!  \sim x^{-C\alpha_s}
\ee
Thus although there is only logarithmic dependence on the
cutoff this can sum up to a power law behavior for the structure
functions.  If this is the case, then it might be that although the
momentum dependence of the structure functions can be determined,
their overall scale is not possible to evaluate.
However, in weak coupling the effect of the overall scale factor
which cannot be computed is of order $A^\alpha \sim 1 + \alpha ln(A)$,
and so long as all the kinematically singular terms have been included,
the undetermined coefficient $A$ which is of order 1 and
depends on the details of
the cutoff only makes a weak coupling perturbative modification
of the structure function.

Higher orders in $\mu^2 $, the infrared structure of the theory is probed.
There are several possibilities which might arise here.  It might turn
out that the infrared singularities of the theory are entirely
screened.  In this case, the infrared properties of all
correlations functions can be computed self-consistently in weak
coupling in the infrared.  A more likely scenario is however that
as in finite temperature QCD, one some of the screening lengths
are essentially non-perturbative.  If this is the case, we can compute
structure functions only in specific kinematic regions, and the behavior
of at least some of the structure functions in the infrared
can be determined only non-perturbatively.

To determine the Feynman rules of the theory, we must evaluate
the propagator.  To zeroth order in $\mu^2$, the propagator is simply the light
cone gauge propagator.  This is
\be
	D^{\mu \nu}_{ab} (k,q) = i\delta_{ab} (2\pi )^4 \delta^{(4)}(k-q)
\left(
g^{\mu \nu} - {{q^\mu n^\nu + n^\mu q^\mu} \over {n \cdot q}}
 +{{q^2 n^\mu n^\nu} \over {(n \cdot q )^2}} \right)
{1 \over q^2}
\ee
where $n$ is the unit vector $n^\mu = \delta^{\mu - }$

To find the first order correction to the propagator arising from
$\mu^2 $ in our effective Lagrangian it is perhaps most easy
to compute the expectation value of
\be
	\Delta D^{\mu \nu}_{ab} (x,y) = \delta <A^\mu_a(x)A^\nu_b(y)>
\ee
by computing with the source present and then integrating out the
source.  We see that this term is given in the presence of the sources
to lowest order as
\be
	\Delta D^{\mu \nu }_{ab}(x,y) = g^2
\delta^{\mu i} \delta^{\nu j}
<{\nabla^i_x \over {\partial^+_x \nabla_t^2}}\rho_a (x)
{\nabla^j_y \over {\partial^+_y \nabla_t^2}} \rho_b(y) >
\ee
Upon integrating over the sources, we have therefore that
\be
	\Delta D^{\mu \nu }_{ab}(k,q) = g^2 \mu^2 \delta_{ab}
(2\pi ) \delta(k^-) (2\pi )
\delta(q^-) (2\pi )^2 \delta^{(2)} (\vec{k_t} - \vec{q_t} )
\delta^{\mu i} \delta^{\nu j} {{k^i_t q^j_t} \over {(k^+q^+ k_t^2 q_t^2)}}
\ee

The change in the propagator has therefore been replaced by
a product of two classical fields times an overall
Kronecker delta function for color invariance and a two dimensional
delta function for invariance under transverse translations.
There is also no factor of $i$ in the change of the propagator.
These are contributions to the dispersive part of the
propagator.  This is always the case in many body theory, where the
dispersive part represents occupied states.

It is now straightforward to compute the distribution function for
gluons.  This may be done by for example computing the expectation
value of the light cone current density.  It may also be read
off directly from the above equation in analogy to the result for
QED.  The result is that

\be
     {1 \over {\pi R^2}} {{dN} \over {d^3k}} =
{{\alpha_s \mu^2 (N_c^2-1)} \over \pi^2}
{1 \over {k^+k_t^2}}
\ee
This is just the Weiszacker-Williams distribution.
The strength of the gluon distribution is proportional to the
typical charge squared per unit area, and should go like
$A^{1/3}$

\section{Summary and Conclusions}

In the previous sections, we discussed the potential problems with
computing the distribution functions in higher order in weak coupling.
Much must be done to turn this hypothetical program into a reality.
Some of the problems to be solved are associated with the ultraviolet.
There are two type of modifications here.  The first are
corrections which generate the proper scaling behavior of
the distribution functions in order to be consistent with
the Altarelli-Parisi equations.  This is presumably straightforward
to do. Following Brodsky, we must introduce a transverse
momentum cutoff, and then require that the theory have the
correct renormalization group improved dependence on this
cutoff.  A second problem is more serious, and that is how to get
any non-trivial behavior in $x_{Bj}$.  This presumably arises from
a dependence on the cutoff which regularizes the delta function
source for the static light cone sources of charge.  The resulting
non-trivial scaling dependence of the structure functions must in
some way be related to the Lipatov Pomeron~\cite{lipatov}.
Needless to say, establishing whether this may be done is entirely
non-trivial and the success of this theory of the structure
functions hinges crucially on being able to establish this fact.
In the end, if the small x behavior is enhanced, the
usefulness of
weak coupling expansion might even be extended to small $A$ targets.
This would make the above formalism much more attractive.

The properties of the infrared structure of this theory
are also entirely non-trivial.  Is the infrared structure
computable in weak coupling, or is it like finite temperature
field theory where some aspect of the infrared structure of the
theory are intrinsically non-perturbative?  Is there a
hierarchy of different infrared scales as there is in
finite temperature theory where the thermal
wavelength is much less than an electric screening length which in turn
is much less than a non-perturbative magnetic screening length?
Are there non-trivial non-perturbative phenomena which occur at
these non-perturbative length scales?

Finally, there remains the problem of computing the quark
distribution functions.  To compute them in lowest order
demands a one loop computation of the fermion propagator.
We intend to do this computation in a later paper.
Of interest here is the ratio of the quark sea distribution
to the small x gluon distribution.  This should be
computable in the same kinematic region where the
gluon distribution is computed.

Finally, there is the issue of how the distribution functions
for quarks and gluons described above are related to the
distribution functions measured in deep
inelastic scattering.  Although at first sight this seems
trivial, recall that we are in the limit where there is a high
density of partons, where one might expect screening effects to
be important.  Moreover, the dependence of the structure
functions on the $Q^2$ of the probe must be established through
use of the Altarelli-Parisi equations.

In all of our analysis in QCD, we never explicitly computed a hadronic
wavefunction.  Is there anyway in QCD to establish what is this
wavefunction, or is it as difficult and in the end unrewarding task
as determining what the wavefunction is for a large system
in the microcanonical ensemble?  In this latter case,
it is sufficient to consider density matrices to study
properties of the system.  The actual wavefunction is almost never
useful.

There is also the question of the relationship between the
distribution functions we have computed and the early time
behavior in heavy ion collisions.  How does one let
the distributions thermalize and evolve through a quark-gluon
plasma?

\section{Acknowledgments}

The authors especially thank Al Mueller for his comments
and help in reading this manuscript prior to
being issued as a preprint, and for guiding us through
the current state of knowledge on this problem.
We wish to thank Stan Brodsky for his help in
understanding the light cone gauge formalism and providing
the authors with extremely clear pedagogical lectures.
We thank Joe Kapusta for encouragement, and Mikhail
Voloshin for critical comments.  We acknowledge support
under DOE High Energy DE-AC02-83ER40105 and DOE Nuclear DE-FG02-87ER-40328.
Larry McLerran wishes to thank the Aspen Center for Physics where part
of this work was finished, and Klaus Kinder-Geiger who in his talk at
Quark Matter 93 implied that to understand the possible formation
of a quark-gluon plasma in heavy ion collisions, one must first understand
the initial conditions.

\end{document}